\documentclass[pre,twocolumn,groupedaddress,showkeys,showpacs]{revtex4-1}
\usepackage{graphicx}
\usepackage{CJK}
\begin{document}
%\begin{CJK*}{GBK}{}

% Use the \preprint command to place your local institutional report
% number in the upper righthand corner of the title page in preprint mode.
% Multiple \preprint commands are allowed.
% Use the 'preprintnumbers' class option to override journal defaults
% to display numbers if necessary
%\preprint{}

%Title of paper
\title{Dense packings of spheres in cylinders  I. Simulations}
\date{\today}
\author{A. Mughal}
\affiliation{Institute of Mathematics and Physics, Aberystwyth University, Penglais, Aberystwyth, Ceredigion, Wales, SY23 3BZ}
\author{H. K. Chan}
\affiliation{Foams and Complex Systems, School of Physics, Trinity College Dublin, Dublin 2, Ireland}
\author{D. Weaire}
\affiliation{Foams and Complex Systems, School of Physics, Trinity College Dublin, Dublin 2, Ireland}
\author{S. Hutzler}
\affiliation{Foams and Complex Systems, School of Physics, Trinity College Dublin, Dublin 2, Ireland}

%\homepage[]{}

%\thanks{}
%\altaffiliation{}

%Collaboration name if desired (requires use of superscriptaddress
%option in \documentclass). \noaffiliation is required (may also be
%used with the \author command).
%\collaboration can be followed by \email, \homepage, \thanks as well.
%\collaboration{}
%\noaffiliation

\date{\today}
\begin{abstract}
  We study the optimal packing of hard spheres in an infinitely long cylinder, using simulated annealing, and compare our results with the analogous problem of packing disks on the unrolled surface of a cylinder. The densest structures are described and tabulated in detail up to $D/d=2.873$ (ratio of cylinder and sphere diameters). This extends previous computations into the range of structures which include internal spheres that are not in contact with the cylinder.
\end{abstract}

% insert suggested PACS numbers in braces on next line
\pacs{}
% insert suggested keywords - APS authors don't need to do this

%\maketitle must follow title, authors, abstract, \pacs, and \keywords

% body of paper here - Use proper section commands
% References should be done using the \cite, \ref, and \label commands

%\maketitle must follow title, authors, abstract, \pacs, and \keywords
\maketitle
%\end{CJK*}
\section{Introduction}

\begin{figure}
\begin{center}
\includegraphics[width=0.5\columnwidth]{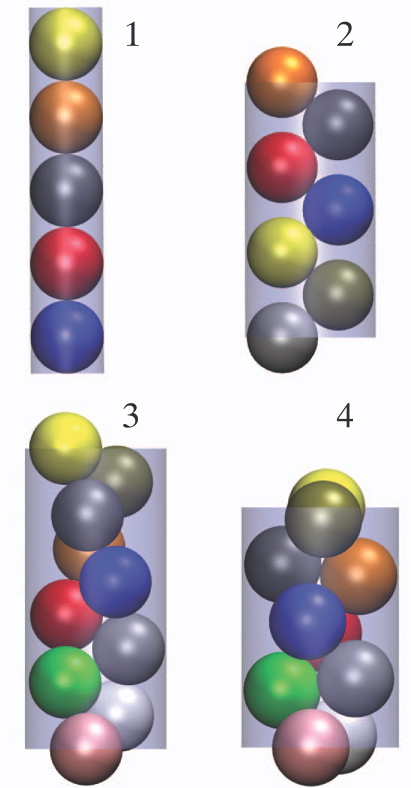}\\
\caption{The first four densest sphere packings in a cylinder.}
\label{simple_structures}
\end{center}
\end{figure}

The dense packing of monodisperse (equal-sized) hard spheres in a cylinder has been found to produce a remarkable sequence of interesting structures as the ratio of the cylinder diameter to the sphere diameter is varied. We have explored computationally the densest of these packings in detail using simulated annealing, following the work of Pickett {\it et al}  \cite{Pickett:2000}, and the first four packings in this sequence are shown in Fig \ref{simple_structures}. We have reported extensive results, together with analytic  approximations which serve to elucidate our findings \cite{Mughal:2011}. The present paper and its future continuations amplify the previous reports by Mughal et al. \cite{Mughal:2011} and by Chan (second author of the present paper) \cite{chan:2011}, and extend them in various directions. In particular, new results are provided for larger cylinder diameters and we include full details of all structures in a form suitable for future reference.

We shall refer to these quasi 1D packings as {\it columnar crystals} since they are periodic. Each structure can be assembled by stacking unit cells \textit{ad infinitum} along the length of the cylinder with each subsequent unit cell rotated by the same twist angle with respect to the previous one.

For smaller cylinder diameters, up to and beyond the range investigated by Pickett et al. \cite{Pickett:2000}, the densest packings only consist of spheres that are in contact with the surface of the cylinder. Eventually at larger cylinder diameters dense packings are found for which there are internal spheres that are not in contact with the surface. Nevertheless we extend the study of structures with only cylinder-touching spheres to larger diameters, for its own sake. As already explained in \cite{Mughal:2011}, these structures are readily understood by recourse to a yet simpler problem, in which circular disks are placed on a cylinder. This can be fully worked out in analytical terms.

In parallel with our computational study using simulated annealing and a separate investigation by Chan \cite{chan:2011} using sequential deposition, a number of experiments have been undertaken on the packing of solid spheres and small bubbles under gravity. Many of the simulated structures are observed. While we will have cause to mention these experiments, their presentation will be reserved for a further paper \cite{Meagher:2012}.

Some of the structures that we describe (particularly the more elementary ones) have been observed in biological microstructures \cite{Erickson:1973}. Columnar crystals have also been realised in numerous experimental contexts including foams (both dry \cite{Pittet:1996, Weaire:1992, Pittet:1995, Boltenhagen:1998, Hutzler:2002} and wet \cite{Meagher:2012}) in tubes, colloids in micro channels \cite{Moon:2004, Moon:2005, Li:2005, Tymczenko:2008, Lohr:2010}, and  fullerenes in nanotubes \cite{Khlobystov:2004, Yamazaki:2008, Warner:2010}.

We have not attempted to rigorously prove that the densest packings identified by simulations are indeed the densest possible. This ought to be achievable for some of the simplest ones.

\section{Motivation and Background}

The identification of dense packings has always played an important part in condensed matter physics and physical chemistry \cite{Aste:2008}. The entities which are to be packed are often spherical, or well approximated as such, and they may be monodisperse.  In the idealised case of hard spheres (with no interaction when not in contact) the problem of finding the maximum density with given boundary conditions constitutes a classic mathematical challenge. If posed for an unbounded packing, it is long associated with the name of Kepler. It is still debated, in terms of full mathematical rigour, although there is no room for practical doubt about the nature of densest packings (fcc etc) \cite{Hales}.

Here we are concerned with the case in which spheres of diameter $d$ are to be contained in an unbounded cylinder of diameter $D$. We shall pursue it up to $D/d = 2.873$ which represents the limit of the capability of our present computational resources and methods.

Our primary objective is to identify the structure with the largest value of $\Phi$, defined as the fraction of volume occupied by the spheres.

There are close connections between this topic and that of \textit{phyllotaxis}, a subject arising out of biology and having to do with the dense arrangement of similar units on the surface of a cylinder, exemplified by pine cones, pineapples or corn cobs  \cite{Airy:1872}. This connection will be explored in detail here.

\section{Some simple columnar crystals}

\begin{figure}
\begin{center}
\includegraphics[width=0.8\columnwidth]{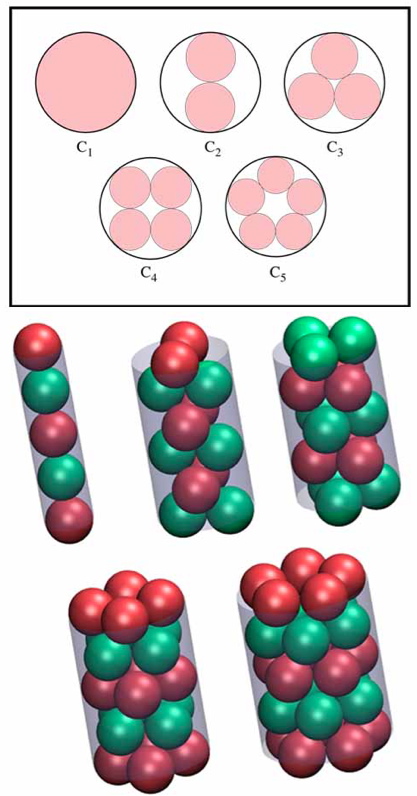}\\
\caption{{\bf Top}: The first five solutions of the circle packing problem. {\bf Bottom}: Five columnar structures corresponding to the first five solutions of the circle packing problem. The unit cells are shown in red/green.}
\label{achiral_states}
\end{center}
\end{figure}

For certain specific values of $D/d$ the optimal packing structure can be easily surmised. These structures are directly related to the analogous two-dimensional problem of finding the smallest diameter circle into which $N$ non-overlapping circles, each of diameter $d$, can be packed. A description of these special cases will serve to illustrate the general problem that we will address.

The first five of these special solutions, labelled $C_N$, are shown schematically in Fig \ref{achiral_states} and (for $N>2$) consist of disks placed at the vertices of a regular polygon. The diameter of the enclosing circle is given by,
\begin{equation}
\
D^c(N)
=
\left\{
\begin{array}{l l}
d & \quad \mbox{if $N=1$}\\
d \left(1+\frac{1}{\sin\left(\pi/N \right) } \right)  %& \quad \mbox{if $2\geq n \geq 4$} \\
 & \quad \mbox{if $ N \geq 2$} \\
\end{array}
\right.
\end{equation}
where $D^c(2)=2d$, $D^c(3)=2.1547d$, $D^c(4)=2.4142d$, $D^c(5)=2.7013d$.

Replacing the circles with $N$ spheres we define a unit cell which can be repeated along the length of the tube as follows. Each successive layer, indicated by alternating red and green spheres, can be generated by translating the previous layer through a distance
\begin{equation}
L^{c}(N)
=
\left\{
\begin{array}{l l}
d & \quad \mbox{if $N=1$}\\
\frac{d}{2}\sqrt{3 - \frac{(1-\cos(\pi/N))}{(1+cos(\pi/N))} } & \quad \mbox{if $ N \geq 2$} \\
\end{array}
\right.
\end{equation}
along the tube and rotating it by a twist angle $\alpha^{c}(N)=\pi/N$. Using the above formula the volume fraction $\Phi$ of these simple  columnar crystals can easily be computed,  $\Phi(1)=2/3$, $\Phi(2)=0.4714$, $\Phi(3)=0.5276$, $\Phi(4)=0.5441$,  $\Phi(5)=0.5370$.

We shall label these simple columnar crystals with the notation $C_N$ to indicate that their structure can be derived in a simple way from the circle packing problem.

\section{Simulation}

\begin{figure*}
\begin{center}
\includegraphics[width=2.0\columnwidth ]{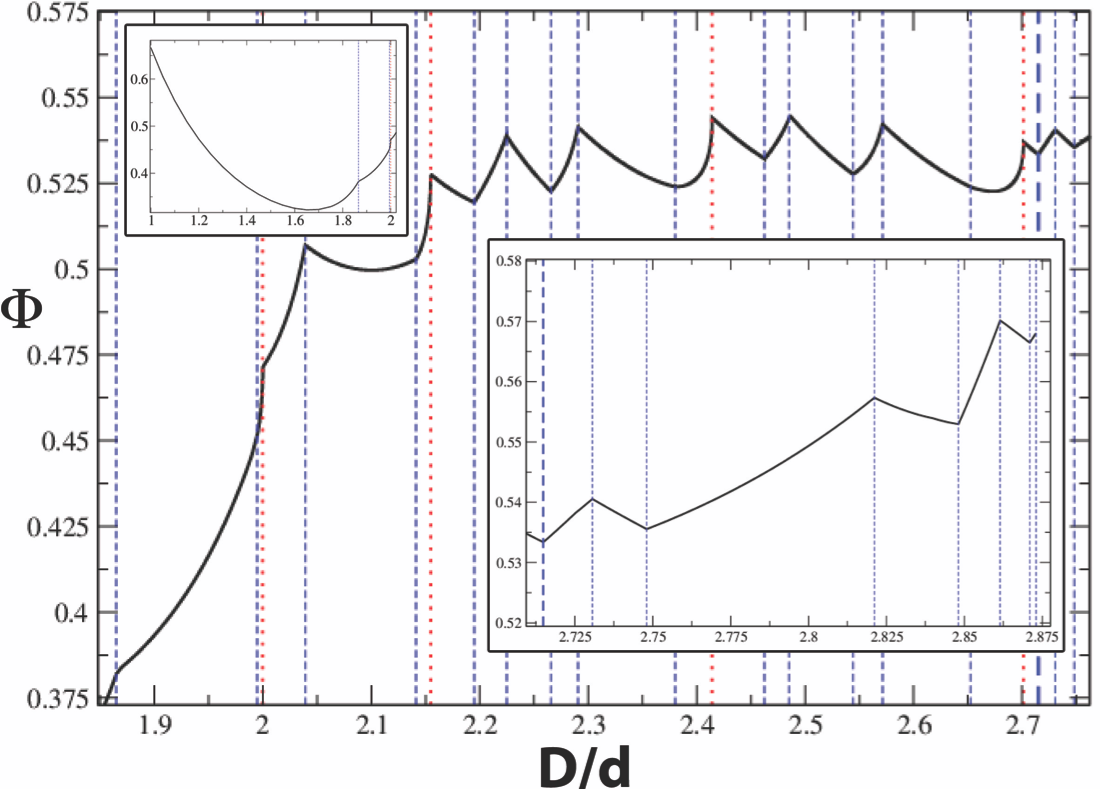}
\caption{Volume fraction of the densest packing as a function of  $D/d$. Discontinuities in the derivative are indicated by vertical lines. Red dotted lines indicate solutions to the circle packing problem. The main graph shows results in the range $1.8<D/d<2.71486$, the inset are continuations of the same graph showing the regions  $1<D/d<1.8$ (left) and $2.71486<D/d<2.873$ (right). Structures above $D>2.71486$ include internal spheres while those below do not, the division between these two regions is denoted by a heavy blue dashed line. }
\label{volume_fraction}
\end{center}
\end{figure*}

For general values of $D/d$, the optimal packing structure cannot be guessed easily and we must turn to heuristic methods. Our search method is confined to structures that are periodic in the following sense.
There is a primitive cell, of length $L$, containing $N$ spheres, the structure being generated from this by the screw operation of (i) translation along the cylinder axis by $nL$  (where $n$ is any integer) combined with (ii) rotation about the axis by an angle  $n\alpha$. This screw operation represents the underlying symmetry of columnar crystals (which are not to be confused with columnar phases, originating in the study of liquid crystals and related instead to the packing of columns).

Our primary method of simulation is based on the well known approach of simulated annealing. This provides a reasonably exhaustive and unbiased  search for maximum density.  Appendix A summarises the technical details of the present application (such as annealing schedules).The search procedure described in Appendix A  looks for the minimum possible value of $L$,  for  a given $N$, treating $\alpha$ and the sphere positions as variables. It does this for increasing values of $N$ until the (likely) optimal structure is apparent, or further computation is not practical.

In the previous simulations of Pickett et al. \cite{Pickett:2000},  a periodic  boundary condition without any twist was used. If for example the densest structure has a twist angle $\alpha$ equal to $\pi/5$, then ten cells would combine to satisfy the simpler boundary condition.

\begin{figure}
\begin{center}
\includegraphics[width=1.0\columnwidth]{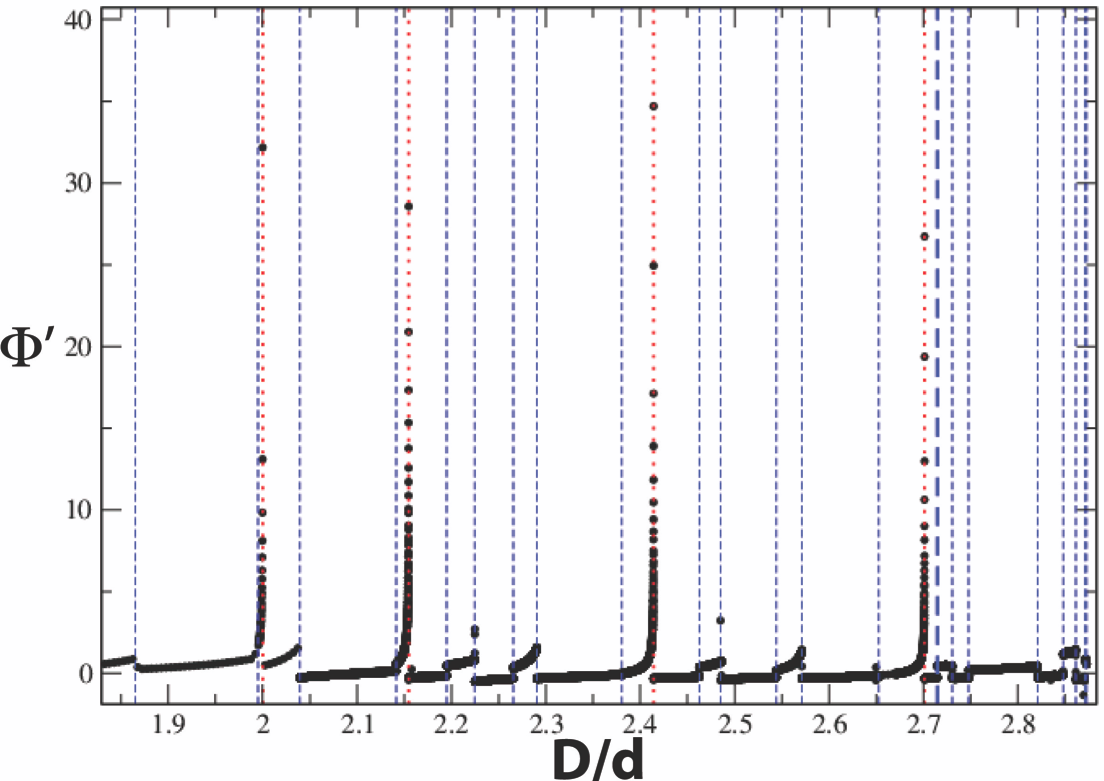}\\
\caption{The numerically computed derivative (finite difference) of the volume fraction curve, as shown in Fig \ref{volume_fraction} }
\label{derivatives}
\end{center}
\end{figure}

\begin{figure}
\begin{center}
\includegraphics[width=0.5\columnwidth]{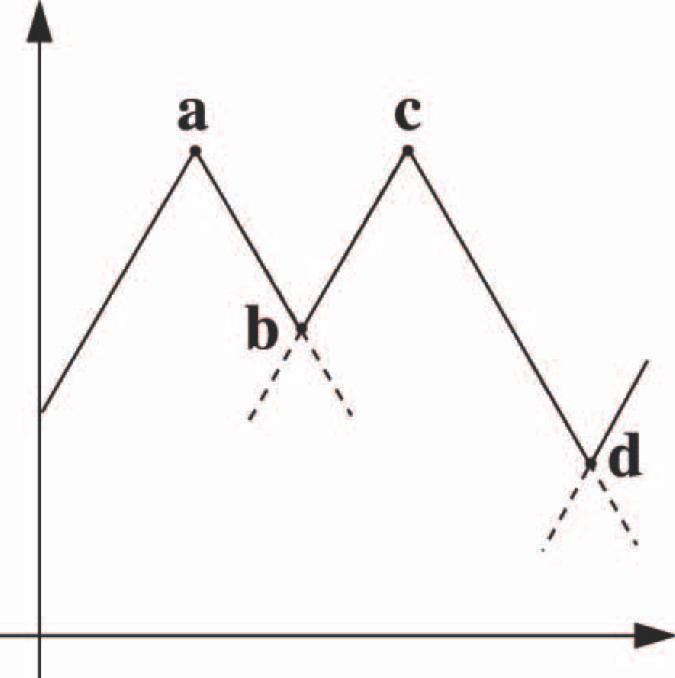}\\
\caption{A schematic representation illustrating variations in the volume fraction of the densest packing arrangements as a function of $D/d$ (see text)}
\label{zigzag}
\end{center}
\end{figure}

\section{Numerical Results}

In this section we present the full range of results for several significant properties as a function of $D/d$. These are the volume fraction and chirality, for the densest structures that we have found. The following figures and Table \ref{exp:vf_table} summarise the main results.

Table \ref{exp:vf_table} enumerates all the densest packings observed, from our numerical work, in the range $1\leq D/d \leq 2.873$. The table lists the range of $D/d$ over which each structure is observed, the number of spheres in the primitive unit cell (from which the extended structure can be constructed) and the average number of contacts per sphere. The table also classifies the structures into two groups: those which are chiral and those which are not (achiral) - see below for a more detailed discussion of chirality.

The break in Table \ref{exp:vf_table}, between structures 32 and 33, highlights the fact that beyond $D/d=2.7379$ the densest structures are those which contain spheres not in contact with the cylindrical boundary. The unit cell for packings with internal spheres are described using the notation $(m,n)$, where the first integer is the number of internal spheres  and the second is the number of spheres which are in contact with the cylindrical boundary. So for example in the case of structure 30 the unit cell contains 6 spheres of which 5 are in contact with the cylindrical boundary and 1 is not.

\begin{center}
\begin{table*}[ht]
 \begin{tabular}{ | l || p{4.0cm} |  p{1.5cm}| p{2.15cm}| p{1.5cm}| p{3.0cm}| p{2.5cm}|}
 \hline
 Structure & Range & Number of spheres in Unit Cell & Notation & Average Contact Number & Description & chirality\\ \hline
 1 ($C_1$) & $ D/d =1$ & 1 &  straight chain & 2 & - & achiral\\ \hline
 2 & $\;\;\;\;1\leq D/d \leq 1.866$ & 1 &  zigzag & 2 & - & achiral\\ \hline
 3 & $\;\;\;\;1.866\leq D/d \leq 1.995$ & 1  & twisted zigzag & 4 & - & chiral\\ \hline
 4 & $\;\;\;\;1.995\leq D/d < 2.0$ & 2  &  ({\bf 2},1,1) & 4 & line-slip & chiral\\ \hline
 5  ($C_2$)& $D/d=2.0$ & 2  & (2,2,0) & 5 & maximal contact & achiral\\ \hline
 6 & $\;\;\;\;2.0 < D/d \leq 2.039$ & 2  & (2,{\bf 2},0) & 5 & line-slip & chiral\\ \hline
 7 & $D/d = 2.039$ & 1  & (3,2,1) & 6 & maximal contact & chiral\\ \hline
 8 & $\;\;\;\;2.039 \leq D/d \leq 2.1413$ & 2  & (3,{\bf 2},1) & 5 & line-slip & chiral\\ \hline
 9 & $\;\;\;\;2.1413\leq D/d < 2.1545$ & 3  & ({\bf 3},2,1) & 16/3 & line-slip & chiral\\ \hline
 10  ($C_3$)& $D/d=2.1547$ & 3  & (3,3,0) & 6 & maximal contact & achiral\\ \hline
 11 & $\;\;\;\;2.1547< D/d \leq 2.1949$ & 3  & (3,{\bf 3},0) & 16/3 & line-slip & chiral\\ \hline
 12 & $\;\;\;\;2.1949\leq D/d \leq 2.2247$ & 2  & (3,{\bf 2},1) & 5 & line-slip & chiral\\ \hline
 13 & $D/d = 2.2247$ & 2  & (4,2,2) & 6 & maximal contact & achiral\\ \hline
14 & $\;\;\;\;2.2247\leq D/d \leq 2.2655$ & 2  & (4,2,{\bf 2})\textbackslash(4,{\bf 2},2) & 5 & line-slip & chiral \\ \hline
15 & $\;\;\;\;2.2655\leq D/d \leq 2.2905$ & 3  & (3,{\bf 3},0) & 16/3 & line-slip & chiral\\ \hline
16 & $D/d = 2.2905$ & 1  & (4,3,1) & 6 & maximal contact & chiral\\ \hline
17 & $\;\;\;\;2.2905\leq D/d \leq 2.3804$ & 3  & (4,{\bf 3},1) & 16/3 & line-slip & chiral\\ \hline
18 & $\;\;\;\;2.3804\leq D/d < 2.413$ & 4  & ({\bf 4},3,1) & 11/2 & line-slip & chiral\\ \hline
19 ($C_4$) & $D/d=2.4142$ & 4  &  (4,4,0) & 6 & maximal contact & achiral\\ \hline
20 & $\;\;\;\;2.4142 < D/d \leq 2.4626$ & 4  &  (4,{\bf 4},0) & 11/2 & line-slip & chiral\\ \hline
21 & $\;\;\;\;2.4626 \leq D/d \leq 2.4863$ & 3  & (4,{\bf 3},1) & 16/3 & line slip & chiral\\ \hline
22 & $D = 2.4863$ & 1  & (5,3,2) & 6 & maximal contact & chiral\\ \hline
23 & $\;\;\;\;2.4863 \leq D/d \leq 2.5443$ & 3  & (5,{\bf 3},2) & 16/3 & line-slip & chiral\\ \hline
24 & $\;\;\;\;2.5443 \leq D/d \leq 2.5712$ & 4  & (4,{\bf 4},0) & 11/2 & line-slip & chiral\\ \hline
25 & $D=2.5712$ & 1  &  (5,4,1) & 6 & maximal contact & chiral\\ \hline
26 & $\;\;\;\;2.5712 \leq D/d \leq 2.655$ & 4  & (5,{\bf 4},1) & 11/2 & line-slip & chiral\\ \hline
27 & $\;\;\;\;2.655 \leq D/d < 2.7013$ & 5  &  ({\bf 5},4,1) & 28/5 & line-slip & chiral\\ \hline
28 ($C_5$)& $D/d = 2.7013$ & 5  & (5,5,0) & 6 & maximal contact & achiral\\ \hline
29 & $\;\;\;\;2.7013 < D/d \leq 2.71486$ & 5  &  (5,{\bf 5},0) & 28/5 & line-slip & chiral\\ \hline  \hline \hline
30 & $\;\;\;\;2.71486 \leq D/d < 2.7306$ & 6 (1,5) &   & 26/6 & -(1,5) & chiral\\ \hline
31 & $ D/d = 2.7306$ & 6 (1,5) &  & 40/6 & maximal contact & chiral\\ \hline
32 & $\;\;\;\;2.7306 < D/d \leq 2.74804$ & 6 (1,5) &  & 26/6 & +(1,5) & chiral\\ \hline
33 & $\;\;\;\;2.74804 \leq D/d < 2.8211$ & 11 (1,10) & & 40/11 &  -(1,10) & see Sect. VIII \\ \hline
34 & $D/d = 2.8211$ & 11 (1,10)  & & 60/11 & maximal contact & achiral\\ \hline
35 & $\;\;\;\;2.8211 \leq D/d < 2.8481$ & 11 (1,10)  & & 58/11 & +(1,10) & chiral\\ \hline
36 & $\;\;\;\;2.8481 \leq D/d < 2.8615$ & 7 (2,5) & & 32/7 & -(2,5) & chiral\\ \hline
37 & $D/d = 2.8615$ & 7 (2,5) & & 34/7 & maximal contact & chiral\\ \hline
38 & $\;\;\;\;2.8615 \leq D/d \leq 2.8711$ & 7 (2,5) & & 30/7 & +(2,5) & chiral\\ \hline
39 & $\;\;\;\;2.8711 \leq D/d \leq 2.873$ & 15 (2,13) & & 72/15 & -(2,13) & chiral\\ \hline
40 & $D/d = 2.873$ & 15 (2,13) & & 90/15 & maximal contact & chiral\\ \hline

\end{tabular}
\caption{Specification of densest structures, up to D/d = 2.873. Bold numerals designate the line-slip type (see text section VII.A). The break in the table denotes the point beyond which packings with internal spheres are observed.}
\label{exp:vf_table}
\end{table*}
 \end{center}

\subsection{Volume fraction}

Fig \ref{volume_fraction} presents the maximum density found by our prescribed procedure, for the full range of $D/d$ that we have explored. By simple finite difference we approximate the derivative of the volume fraction as a function of $D/d$, as shown in Fig \ref{derivatives}. This is to clearly identify the singular behaviour discussed below.

The vertical lines indicate a discontinuity in the derivative of the volume fraction; there is either (i) only a sudden changes in the derivative, or (ii) the simultaneous presence of such a sudden change and a square-root singularity in the derivative, the two cases of which are denoted by vertical blue-dashed and red-dotted lines, respectively. Note the square-root singularities in the derivative coincide with the $C_N$ (circle packing) structures.

Let us discuss this behaviour by reference to Fig (\ref{zigzag}) which is a simplified cartoon of the observed changes in the volume fraction as a function of $D/d$. Qualitatively, the variation of the maximum density takes the following form: its dependence on $D/d$ is everywhere continuous while the derivative is piecewise continuous. We distinguish two types of singular points, as in Fig (\ref{volume_fraction}), at which the derivative changes.

The first occur when a maximum number of contacts is reached. These points correspond to highly symmetric structures, such as the $C_n$ structures described above.  At such points - represented by the vertices labelled $a$ and $c$, in the inset - the previous trend of structural change cannot be continued; in effect, the structure has {\it jammed} due to the formation of new contacts. In other words: if the separations of contacting sphere centres is to be maintained, the system is overconstrained and no such solution exists beyond this point. Instead, some existing contacts are released and a new structural trend proceeds. The structure itself varies continuously through the singular point, with a downward change in the derivative of the density, corresponding to the line segments $ab$ and $cd$ in the cartoon.

At the second kind of singular point the structure is simply overtaken by another more dense packing.  Here the structure itself changes discontinuously, and the derivative obviously must change in a positive sense. In the inset this transition corresponds to the points labelled $b$ and $d$ where, for example, the line segment $bc$ illustrates the increasing trend in density immediately following such a transition. The dashed lines indicate the continuation of the previous structure which now has a lower density compared with the optimal packing.

These remarks are based on observation of the results. We have for example no proof that the change of derivative at points of the first type is always negative, although we can offer a proof that the dependence of (maximum) density on $D/d$ cannot have any discontinuities at which the density undergoes a finite upward or downward change. This is discussed in detail in appendix C.

\subsection{Chirality}

\begin{figure*}
\begin{center}
\includegraphics[width=1.8\columnwidth]{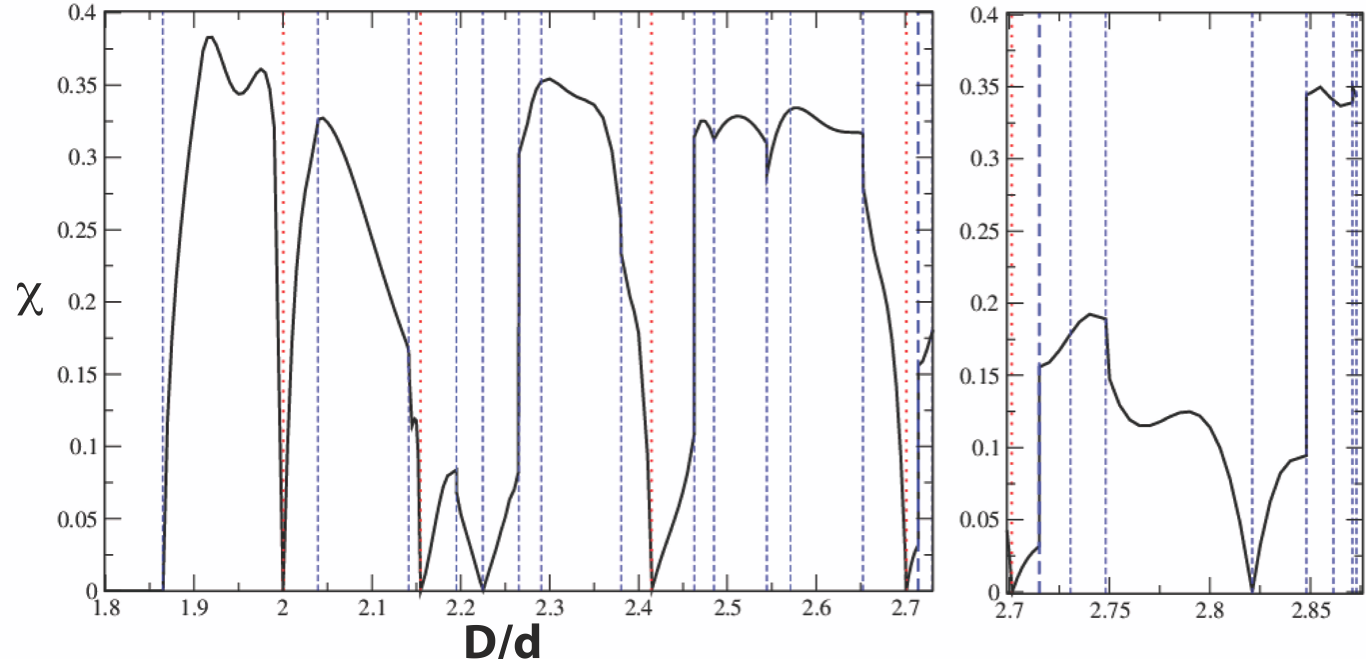}\\
\caption{A plot of the chiral index $\chi(D/d)$, which measures the degree of mismatch between a packing and its mirror image (see text). Vertical lines indicate structural transitions as in Fig 2. The main graph, on the left shows results in the range $1.8<D/d<2.71486$ while the graph on the right is a continuation but focuses on the region $2.71486<D/d<2.873$.}
\label{chirality}
\end{center}
\end{figure*}

A chiral object is one which cannot be superimposed on its mirror image (or inverse)by translations and (proper) rotations. The question of chirality is interesting in the present context, for example in relation to designing chiral molecular filters that will discriminate between enantiomers.

According to this definition an object is either chiral or it is not (it is achiral). The reported structures are classified accordingly in Table \ref{exp:vf_table}.

But chirality may be manifested in physical properties to a greater or lesser extent: the degree of chirality of the structure itself is a tempting concept. While it may be useful to think of this in practice, there is no unique quantitative definition that will have general relevance. {\it Any} appropriate property could be used as an index of chirality. Nevertheless one may offer a working definition for use in relation to simulations, as Pickett {\it et al} did \cite{Pickett:2000}. Here we will employ another definition that is perhaps more transparent.

We define a chirality index $\chi$ in terms of the degree to which an object can be superimposed upon its mirror image. For a given columnar packing, our method is to start with the sphere centres and generate a mirror image of the structure by reflecting the coordinates of the centres in the ${\bf x}$-${\bf y}$ plane, i.e. the cross-sectional plane of the cylinder. For each sphere in the packing we compute its distance to the nearest sphere in the mirror image. The overlap function is defined as the sum of these distances divided by the number of spheres in the original packing. Clearly when a packing can be completely superimposed upon its mirror image the overlap function vanishes.

The computational challenge then is to find the arrangement of the mirror image (by rotation and translation) that minimises the overlap with respect to the original structure. This is done in a straightforward manner by simulated annealing. Clearly for chiral structures the overlap cannot vanish and by plotting the overlap function we have a measure $\chi$ of the chirality as a function of $D/d$, as in Fig(\ref{chirality}). 

\section{Details of Structures for $D/d$ up to 2.71486}

Up to $D/d=2.71486$, all the structures that are found are of a special character, helpful for their interpretation.  Every sphere is in contact with the cylinder surface. Therefore all of their centres lie on an inner cylinder, of diameter D-d, and they touch another cylinder of diameter D-2d. They may therefore be considered as the densest packings of spheres {\it on} a cylinder.

Clearly inner spheres must appear in the optimal packing when D-2d is greater than d, that is $D>3.0$, in fact they are first found at $D/d > 2.71486$, and we reserve them for a separate discussion in the next section. We also treat structures for low $D/d$ separately in section \ref{sec:low_D_structures}.

\subsection{Maximal contact structures}

\begin{figure*}
\begin{center}
\includegraphics[width=1.8\columnwidth]{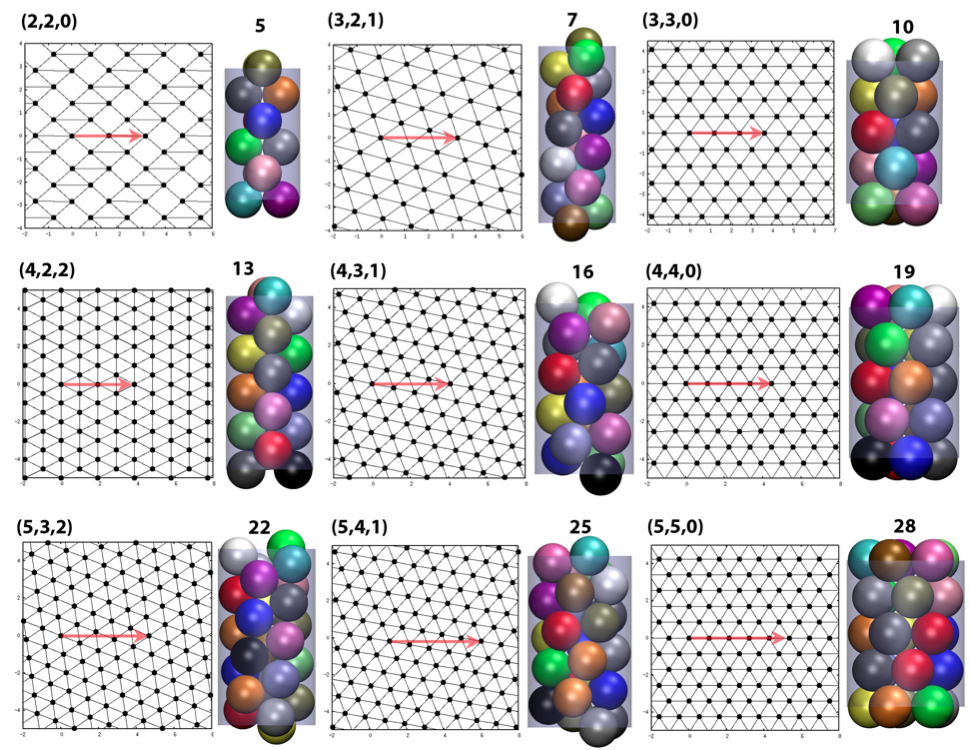}\\
\caption{Maximal contact structures, with the corresponding ``rolled-out'' pattern of contacts. The vector {\bf V}} (corresponding to the perimeter of a cylinder cross-section) is indicated by the red arrows.
\label{composite_symmetric}
\end{center}
\end{figure*}

We have already noted the special points at which the structure changes continuously, with the formation of new contacts and the breaking of old ones. We have previously called them {\it symmetric} structures, but this term is really only meaningful in the disk-packing analysis below. Here we shall apply the term {\it maximal contact} instead. For $D/d\geq2$, all of these structures are depicted in Fig (\ref{composite_symmetric}) and labelled using the phyllotactic notation explained in Appendix B.

They include the simple $C_n$ packings which we have already described and are labelled $n,n,0$ in the {\it phyllotactic notation}, see Appendix B. This relates to the  pattern of sphere centres and contacts, ``rolled out" on to a plane, and identified with rhombic or triangular lattices. This does not apply to the first of the structures, the straight chain, and is rather confusing if applied to the next two, hence we use it only for the subsequent structures.

We begin with the case of $D/d=2$: this is the close-packed, achiral, structure $C_2$ previously described. Each sphere has five contacts (not including contact between a given sphere and the cylinder); these consist of a contact with the neighbouring sphere in the same unit cell and four contacts with spheres in adjacent cells. All of the rest of the maximal contact structures so far considered are composed of spheres with six contacts. Packings of the type $C_n$ are achiral while the rest are chiral.

 \subsection{Line slip structures}

\begin{figure*}
\begin{center}
\includegraphics[width=2.0\columnwidth]{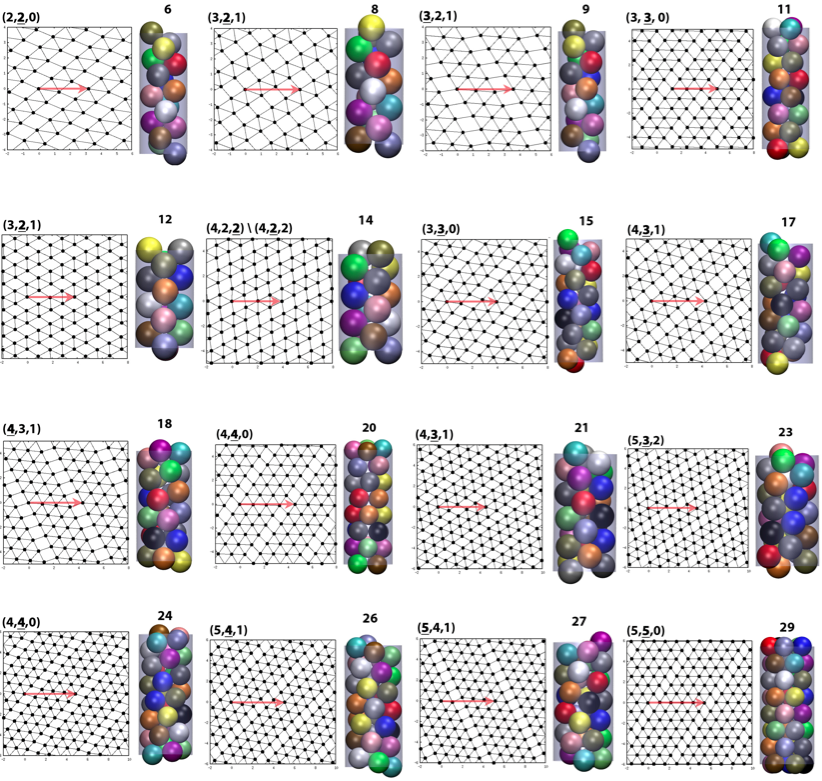}
\caption{Line slip structures at arbitrarily chosen values of $D/d$ within their respective ranges, with corresponding rolled out patterns of contacts. The vector {\bf V} is indicated by an arrow.}
\label{line slip composite_symmetric}
\end{center}
\end{figure*}

We now turn to the intervening structures, which are modified versions of the maximal contact packings, adjusted to fit around the cylinder.  Such {\it staggered helices} were first observed by Pickett {\it et al} in their simulations. We label these {\it line slip} structures to indicate that the modification consists of the release of half of the contacts along one line separating two of the spiral chains of the symmetric structure, and a consequent relative slip of the two sides. Such line slip arrangements are a feature of hard sphere packings and are evident in the rolled out patterns of Fig \ref{line slip composite_symmetric}.

There are in general three possible choices for the direction of the line slip and the results of Fig \ref{line slip composite_symmetric} correspond to those that maintain the highest density as in Fig \ref{volume_fraction}.

We continue to higher values of $D/d$ in section VIII.

\subsection{Structures below $D/d=2.0$, and the square root singularity}
\label{sec:low_D_structures}

We return to examine the structures for low D/d, which are somewhat different. The reason for such a difference will become clearer when we explore the relationship to disk packing, in a later section. These first four structures are shown in  Fig \ref{simple_structures}.

Structure 1 is the elementary case of a straight chain of spheres, like peas in a pod, as shown in Fig \ref{simple_structures}. This trivial, achiral,  structure has a volume fraction of $\Phi(1)=2/3$. Similarly structure 2 is obvious and  consists of a zigzag planar arrangement of spheres such that the change of the azimuthal angle from one ball to the next is equal to $\pi$; each sphere makes contact with one sphere from above and one from below. When the zigzag packing, i.e structure 2, encounters additional contacts between its second neighbours at $D/d= 1.886$ (so that each sphere now has four contacts), it is forced to take the form of an increasingly twisted spiral (structure 3). By direct numerical calculation one may follow this to its end point at structure 7, but structure 6 intervenes with higher density, so that there must be a transition below D/d = 2.0. This is found at D/d=1.990, strikingly close to 2.0, beyond which the densest packing is a line-slip modification of structure 6. The smallness of the interval in which this is found is largely due to the existence of a square-root singularity.

The variation of volume fraction, as any of the $C_n$ or $n,n,0$ structures is approached with increasing $D/d$, exhibits a square-root form, whereas it is linear in other cases. That is
\begin{equation}
\Phi_0-\Phi
\approx
\left(
\left(\frac{D_0}{d_0}\right) -
\left(\frac{D}{d} \right)
\right)^{1/2},
\end{equation}
where the quantities with a subscript 0 belong to a $C_n$ or $n,n,0$ structure. The derivative plot shown in Fig 2 is intended to make this singular behaviour more evident. The square-root arises from the special symmetry of these achiral structures.

\section{Relationship to circular disk packings on a cylinder}

\begin{figure*}
\begin{center}
\includegraphics[width=2.0\columnwidth ]{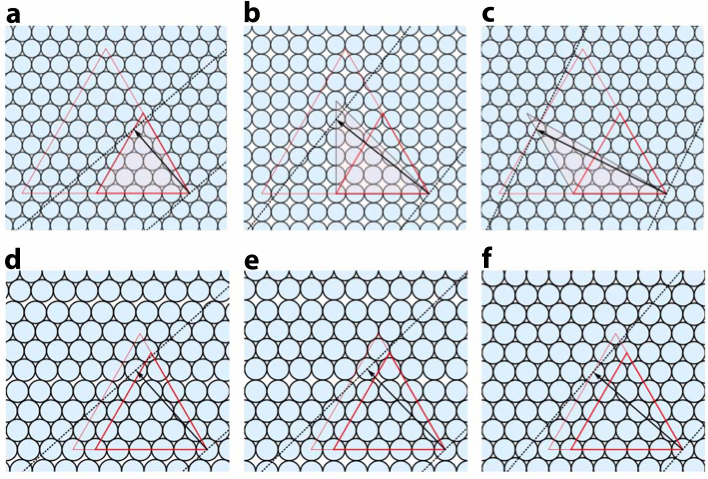}
\caption{(a) symmetric packing [5,4,1], the black arrow represents the periodicity vector; (b) the square packing [5,4]; (c) the symmetric packing  [9,5,4] which is connected by an affine shear to [5,4,1];  (d \& e) show a line slip connecting [5,4,1] to the symmetric packing [6,4,2] (f).}
\label{disk_packing}
\end{center}
\end{figure*}

Many aspects of these results are made more understandable by recourse to the packing of circular disks on a cylinder. We have already noted that the same sequence of ÒsymmetricÓ structures is found. The similarity extends further to the line-slip structures and some of the details of the analytic form of the curve in Fig (\ref{volume_fraction}), not yet noted.

This {\it qualitative} correspondence attracts our attention to the cylindrical disk packing problem, which we pursue below.

\subsection{Disk Packings}

Densest packing of circular disks in a plane places their centres on a triangular lattice where each disk has six contacts; this was rigorously proved about a century ago \cite{Aste:2008}. Cylindrical surface packing of disks with the same density is generated by rolling this pattern on to a cylinder, when possible.

To see how this can be achieved seamlessly let us define a periodicity vector ${\bf V}$ between any pair of disk centres of the triangular lattice packing as shown in Fig (\ref{disk_packing})a. The region between the start and end of ${\bf V}$, which is bounded by lines perpendicular to ${\bf V}$, can then be excised and wrapped onto a cylinder whose circumference is equal to the length of ${\bf V}$. The resulting structure is a dense, homogenous packing - in the sense that all disk sites are equivalent - which we call a symmetric packing. Any such symmetric packing is characterised by this periodicity vector. In the phyllotactic scheme ${\bf V}$ can be defined by a set of three ordered positive integers $(l=m+n,m,n)$, as discussed in appendix B (a simple operational method to assign indices is to count the number of lattice rows that cross the periodicity vector in the three directions, as shown in Fig (\ref{disk_packing})a and Fig (\ref{disk_packing})c).

For other values of the cylinder circumference, the symmetric packing may be distorted in some way to wrap around the cylinder. The most obvious adjustment is an affine transformation of the triangular lattice, as shown in Fig (\ref{disk_packing})b. We illustrate the effect of the transformation by reference to the shaded triangle in Fig (\ref{disk_packing})a. The affine transformation distorts the equilateral triangle into an isosceles triangle by keeping the length of two adjacent sides fixed and varying the third; consequently, the disk centres form a rhombic lattice which has a lower density (area fraction) compared with the triangular lattice since each disk now has only four contacts (the lowest area fraction corresponds to a square lattice).The transformation may be adjusted to make the length of $V$ equal to  $\pi D$. The resulting pattern can be wrapped onto the surface of the cylinder. We may call these asymmetric or affine lattice packings.

If $D/d$ is varied the disk centres of such a structure are eventually brought back into coincidence with the sites of a triangular lattice, see Fig (\ref{disk_packing})c. In this manner the strained structure proceeds from one symmetric packing to another. Since there are three possible choices for the affine transformation, the rules for this process - as reported previously \cite{Mughal:2011} - are, when applied to the second and third phyllotactic indices, as follows,
\begin{eqnarray}
(m,n)
&\rightarrow&
(m-n, n)
\nonumber
\\
(m,n)
&\rightarrow&
(m+n, m)
\nonumber
\\
(m,n)
&\rightarrow&
(m+n, n),
\nonumber
\end{eqnarray}
where the new phyllotactic indices may have to be rearranged into descending order. Intermediate asymmetric packings may be labelled using rhombic notation $[p,q]$, where the ordered indices $p\geq q$ are the indices common to both the initial and final symmetric states.

We illustrate the use of these rules with an example. In the case of the symmetric packing $[5,4,1]$  (illustrated in Fig (\ref{disk_packing})a) the above rules yield,
\begin{eqnarray}
(4,1)
&\rightarrow&
(3, 1)
\nonumber
\\
(4,1)
&\rightarrow&
(5, 4)
\nonumber
\\
(4,1)
&\rightarrow&
(5, 1).
\nonumber
\end{eqnarray}
Thus the symmetric packing $[5,4,1]$ is connected to $[4,3,1]$ via the rhombic structure $[4,1]$, to $[9,5,4]$ via $[5,4]$ - which are shown in Fig (\ref{disk_packing})c and Fig (\ref{disk_packing})b, respectively, and to $[6,5,1]$ via $[5,1]$.

\begin{figure*}
\begin{center}
\includegraphics[width=1.5\columnwidth ]{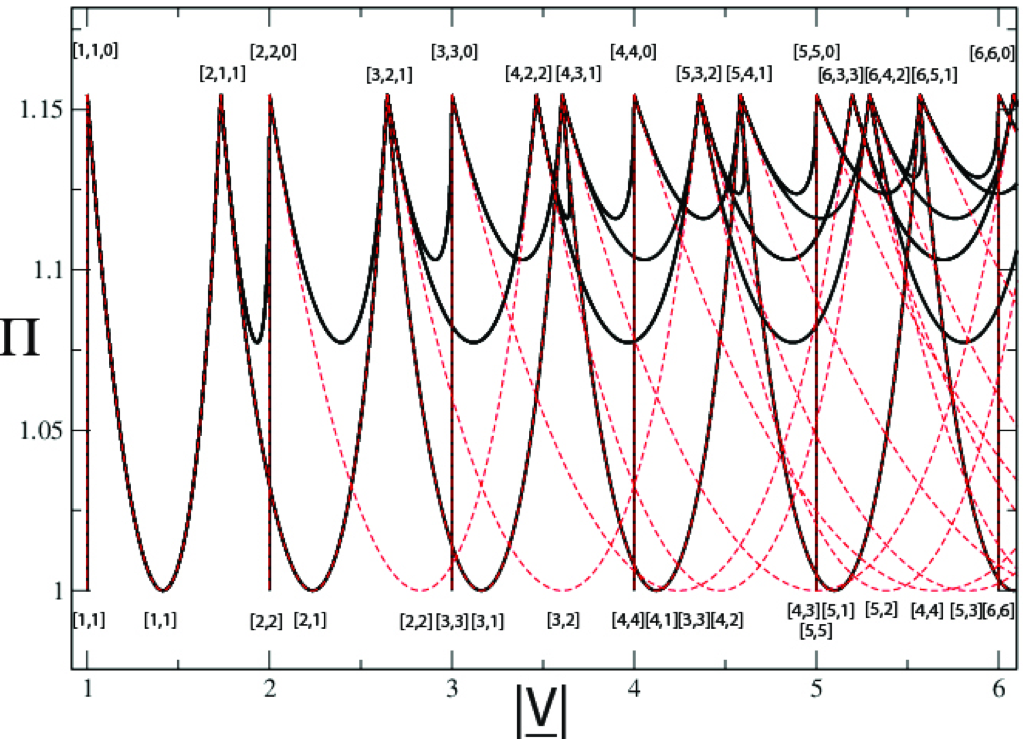}
\caption{Area fraction $\Pi$ of disk packings on the plane which are consistent with wrapping onto a cylinder of diameter $|{\bf V}|/\pi$. Asymmetric packings are shown by (red) dashed curves; line-slip solutions are (black) continuous curves. Symmetric structures labelled $[l,m,n]$ correspond to points of maximum density while intermediate asymmetric packings are labelled using the rhombic notation $[p,q]$}
\label{area_fractions}
\end{center}
\end{figure*}

Although such asymmetric packings are the simplest type of dense structure, intermediate between two symmetric packings, they are not the densest. As reported previously \cite{Mughal:2011}, asymmetric packings are in general superseded by another type of packing involving an inhomogeneous shear of the symmetric lattice in which there is a localised strain or slip along a line (and its periodic replicas).

We can describe such {\it line-slip} packings with reference to the symmetric packing $[5,4,1]$. Again there are three possibilities and we describe each of these in turn. In the first case, four lattice rows cross the periodicity vector in the horizontal direction, as shown in Fig (\ref{disk_packing})a. We allow the final, fourth, row to ``slide-over'' the previous row until the disk centres are once again arranged into a triangular lattice. Thus as illustrated in Fig (\ref{disk_packing}), the line slip in question is intermediate between the the symmetric packing $[5,4,1]$ and $[6,4,2]$. By allowing the disks in the final row to slide in the opposite manner we find that the symmetric packing $[5,4,1]$ is connected to $[4,4,0]$. In the second case five lattice rows cross the periodicity vector. Four rows are held fixed while the final layer of disks slides over the penultimate row - thus $[5,4,1]$ is connected to $[5,3,2]$ and $[5,5,0]$. In the third case only a single lattice row crosses the periodicity vector and we find that  $[5,4,1]$ is connected to $[6,5,1]$ and $[4,3,1]$

This localised shear allows the length of ${\bf V}$ to vary continuously. The disks involved in the line-slip have five contacts while the rest of the structure remains symmetric and close-packed, with six contacts for every disk. Clearly the area fraction has a maximum, corresponding to the symmetric packings, while the intermediate line-slip packings have a lower value. As reported previously \cite{Mughal:2011}, there is a simple rule for the close-packed structures that are the end points of line-slip solutions
\begin{eqnarray}
(1)  \;\;\;\;\;\;\;\;\; (m,n)
&\rightarrow&
(m+1, n) \;\;\;\;\;\;\;\;\;  \textrm{or} \;\;\; (m-1, n)
\nonumber
\\
(2)  \;\;\;\;\;\;\;\;\; (m,n)
&\rightarrow&
(m, n+1) \;\;\;\;\;\;\;\;\;  \textrm{or} \;\;\; (m, n-1)
\nonumber
\\
(3)  \;\;\;\;\;\;\;\;\; (m,n)
&\rightarrow&
(m+1, n-1) \;\;\; \textrm{or} \;\;\; (m-1, n+1),
\nonumber
\end{eqnarray}
where the leading numbers denote the direction $\widehat{\bf a}_1$,  $\widehat{\bf a}_2$ or $\widehat{\bf a}_3$ of the line-slip. Again the above rules apply to the second and third phyllotactic indices of a given close packed structure and keep either $n$, $m$ or $l$ constant. For example, using the above rules, the symmetric packing $[5,3,2]$ is connected by a line slip along $\widehat{\bf a}_1$ to $[6,4,2]$ or $[4,2,2]$, a line slip along $\widehat{\bf a}_2$ yields $[4,3,1]$ or $[6,3,3]$ and a line slip along $\widehat{\bf a}_3$ yields $[5,4,1]$ or $[5,3,2]$ (note in the third case - along $\widehat{\bf a}_3$ - a rearrangement of the phyllotactic indices into descending order was necessary and the new structure is the same as the initial structure). 

In Table \ref{exp:vf_table} we use bold numerals to denote the direction of the line slip. In the table we only have cause to mention the first column of line-slip structures, that is: we denote
\begin{eqnarray}
(1)  \;\;\;\;\;\;\;\;\; [l,m,n]
&\rightarrow&
[l+1,m+1, n] \;\;\;  \textrm{by} \;\;\; [l,m, {\bf n}]
\nonumber
\\
(2)  \;\;\;\;\;\;\;\;\; [l,m,n]
&\rightarrow&
[l+1,m, n+1]  \;\;\; \textrm{by} \;\;\; [l,{\bf m}, n]
\nonumber
\\
(3)  \;\;\;\;\;\;\;\;\; [l,m,n]
&\rightarrow&
[l,m+1, n-1]  \;\;\; \textrm{by} \;\;\; [{\bf l},m, n],
\nonumber
\end{eqnarray}

\begin{figure}[tbp]
\includegraphics[width=7cm]{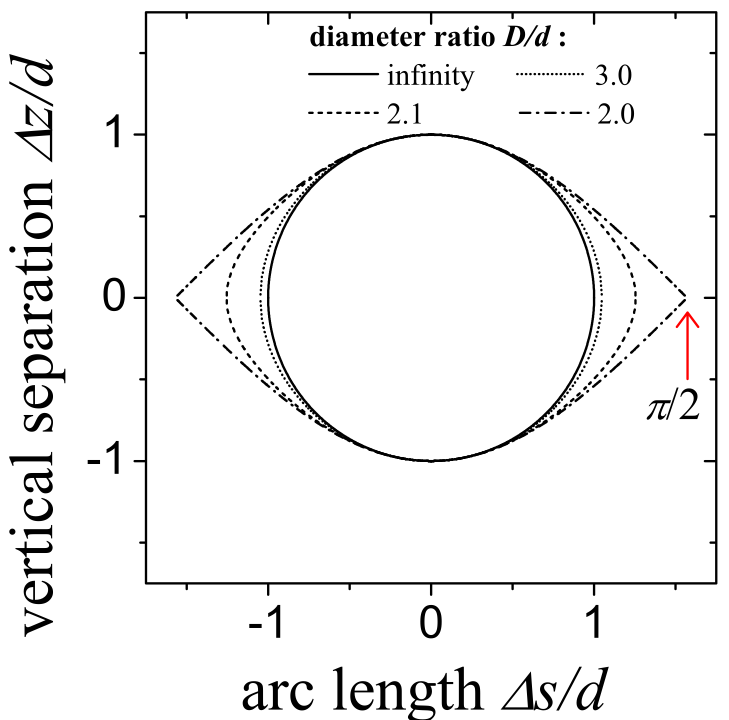}
\caption{A plot of Eq. (\ref{eq:contact}) for various values of $D/d$, showing the approach towards elliptical, and eventually circular, packing as $D/d$ increases beyond 2.}
\label{contact_equation}
\end{figure}

A full derivation of such connection rules for both the asymmetric and line-slip packings will be given in a future publication.

Fig (\ref{area_fractions}) presents analytical results for the density of these disk packings. Here we see for the first time some of the structures of lower density, as well as those of maximum density. Even at this stage, we find a qualitative resemblance between the Fig (\ref{area_fractions}), the disk packing problem, and the corresponding sphere packings (as shown in Fig (\ref{volume_fraction})).

A simple method for achieving a semiquantitative description of the sphere packing problem relies on the fact that for the range $1<D\leq 2.71486$ all the spheres in the packing are in contact with the cylindrical boundary. Thus there exists an {\it inner} cylinder on which the centres of all the spheres are to be found. The pattern formed by the intersection of the spheres with the surface of the inner cylinder bears a close resemblance to the disk packing problem, which we now consider in detail.

\subsection{Relation to sphere packing}

We now explore more closely the {\it quantitative} correspondence between disk and sphere packings, and find a way of bringing them as close to each other as is possible without undue complication.

Consider again a packing of spheres, all touching the confining cylinder of diameter $D$. Their centres lie on an inner cylinder of diameter $D'=D-d$, so that a given centre is located at $\left(D'/2, \theta, z \right)$ in cylindrical polar coordinates. The separation of contacting spheres in 3D is $d$ but if the inner cylinder is rolled out onto the plane, i.e. a given sphere centre is mapped to the 2D cartesian coordinates  $\left(s=D' \theta/ 2, z \right)$, then the distance between the centres of contacting spheres in 2D depends on their mutual orientation.

In general, the required contact rule, in 3D, for a pair of spheres (each having diameter $d$) whose centres, in cylindrical polar coordinates, are located at ${\bf r}_1=(D'/2, \theta, z)$ and ${\bf r}_2=(D'/2, \theta+\Delta\theta, z+\Delta z)$ is \cite{chan:2011},

\begin{equation}
\
(\Delta z)^2
+
\frac{D'^2}{2}
(1-\cos \Delta \theta)
=
d^2,
\label{eq:contact}
\end{equation}
We plot the parametric equation Eq. (\ref{eq:contact}) in Fig. \ref{contact_equation}, for various values of $D/d$ (where the distance $\Delta s\equiv(D'/2)\Delta\theta$ is referred to as the `arc length'). We see that, as $D/d$ increases above 2, the `surface packing' resembles the packing of oval disks, which may be approximated by ellipses. For contacting spheres that lie directly in a line parallel to the direction of the cylinder axis, the separation of their centres is still $S_{||}=d$, while for those that lie in the perpendicular plane, it is,
\begin{equation}
S_{\perp}
=
D'\sin^{-1}
\left(
\frac{d}{D'}
\right).
\end{equation}
For simplicity we take the major and minor axes of the ellipse to have the $S_{\perp}$ and  $S_{||}$, so that the distance on the plane for a pair of contacting spheres is approximated by,
\begin{equation}
S^2 =
Êd^2 \sin^2(\phi)
+ Ê
Ê\left[
D'\sin^{-1}
\left(
\frac{d}{D'}
\right)
\right]^2
\cos^2(\phi).Ê
\end{equation}

Comparing the circumference of the inner cylinder $\pi D'$ with the length of the periodicity vector $Vd$ (measured in units of the disk diameter $d$), gives a stretch factor,
\begin{equation}
X
=
\frac{\pi D'}{Vd}
=
\frac{S_{\perp}}{d},
\label{eq:stretch_factor}
\end{equation}
where the last term is the ratio of the arc distance between contacting spheres, which lie perpendicular to the direction of the cylinder axis. In this way the sphere packing is related to the  planar packing of elliptical disks, which is merely a stretched version of a circular disk packing. Given any packing of circular disks in the plane, it may simply be stretched in the direction of the vector ${\bf V}$, by a factor $X$, and wrapped onto a cylinder of diameter
\begin{equation}
D'=\frac{d}{\sin(\pi/V)}
\end{equation}
(obtained using Eq. ($\!\!$~\ref{eq:stretch_factor})), to create a good approximation to a sphere packing.

\begin{figure*}
\begin{center}
\includegraphics[width=1.5\columnwidth ]{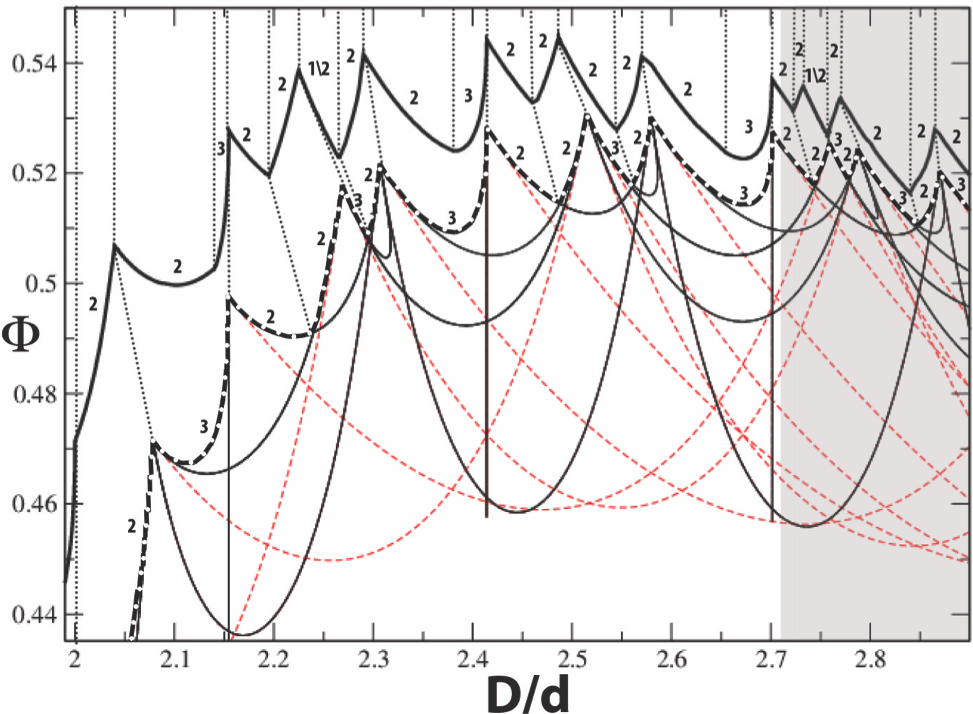}
\caption{Comparison of simulated and analytic volume fractions for the densest packings - upper and lower curves, respectively. Dotted lines are a guide for the eye to the detailed correspondence. In the analytic results, the line-slip structures are identified by black continuous lines, and the red dashed-dotted lines denote asymmetric packing structures. The numbers 1, 2, and 3 (see text) denote the type of line slip observed (upper curve) or predicted (lower curve)Ñwith $1\backslash 2$ denoting a degenerate case. The shaded region on the right is the continuation of the single layered structures for $D/d>2.71486$.}
\label{volume_fractions}
\end{center}
\end{figure*}

Thus a correspondence that at first seems distant may be brought much closer. Fig (\ref{volume_fractions}) presents the resulting transformed curves for sphere packing density; asymmetric and line slip structures are shown by red (dashed) and black (continuous) lines, respectively. The peaks in the density correspond to symmetric packings. The lower (dashed) heavy black line accentuates the curve of maximum density, as predicted by our analytical approach. This is to be compared with the results of simulated annealing shown by the upper curve. We have extended our simulated annealing study of single layer packings to $D>2.71486$, as shown by the grey region in  Fig (\ref{volume_fractions}), by searching for the densest structures in which all the spheres have centres on the surface of a cylinder of diameter $D'$. From this we see that the maximum possible density, for single layer packings, is found at $\Phi(2.4863)=0.5446$ and corresponds to the symmetric packing $(5,3,2)$. Beyond this the packing density steadily diminishes since spheres are to be found only on the surface of the cylinder and the interior is empty.

The analytical method presented here gives the correct line-slip solution in the majority of cases. There are, nevertheless, notable discrepancies between the analytical and numerical results. These are due to the simplicity of the transformation used to map the disk packing problem into the sphere packing problem; as a consequence, the analytical results always predict a type (3) line-slip solutions leading up to $n,n,0$ packing while simulations in fact find a region split between type (2) line-slip followed immediately by a type (3) line-slip. A more accurate (and more complicated) transformation ought to account for this by pushing the type (2) curve above the type (3) solution, for part of this region. This is borne out by the fact that as the diameter of the cylinder increases, and higher order corrections to the transformation diminish in importance, the type (2) solution in the simulations are observed over an increasingly smaller range in the split region.

\section{Structures beyond D/d  = 2.71486, with internal spheres.}

\begin{figure*}
\begin{center}
\includegraphics[width=2.0\columnwidth ]{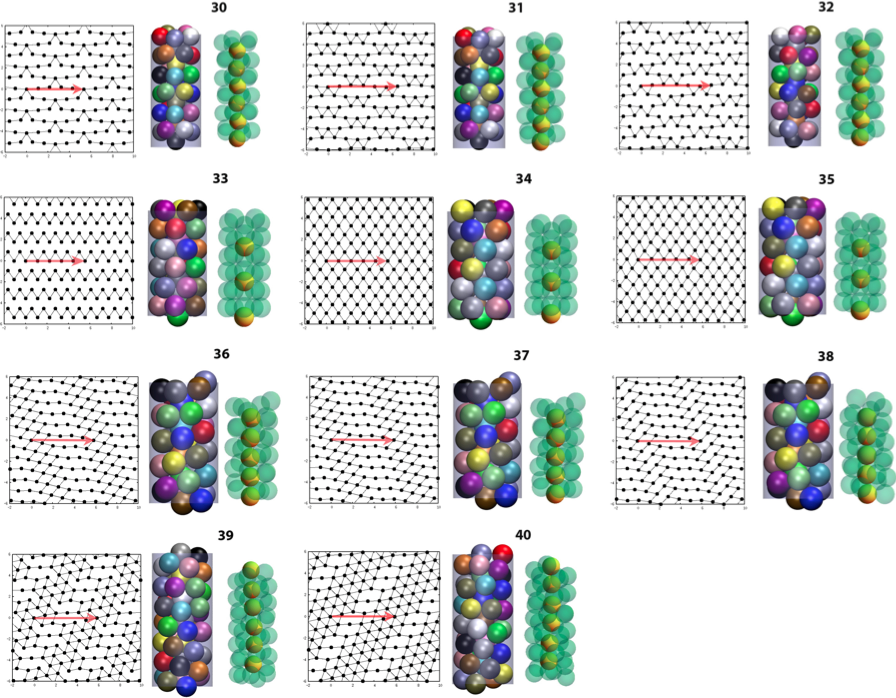}
\caption{Structures with internal spheres. In the case of structures which are not maximal contact the images are produced from numerical results for arbitrarily chosen values of $D/d$ within their ranges. The central image shows each structure viewed side-on. The image on the right shows the inner layer by making the outer layer transparent. The image on the left, for each packing, is the rolled out structure generated by the outer layer of spheres which contact the cylindrical boundary - the periodicity vector $\bf V$ is indicated by the red arrow.}
\label{higher_order}
\end{center}
\end{figure*}

We are able to pursue the same simulations to higher cylinder diameters, finding the anticipated occurrence of internal spheres, initially at $D/d =2.71486$. The structures involved are tabulated in the final section of Table \ref{exp:vf_table}. In some cases phyllotactic notation may still be useful, but much of the simplicity of the previous sections is lost, and increasingly so. The extension of the curve of volume fraction, as far as present computing resources allow, is shown on the right hand side of Fig (\ref{volume_fraction}).

It is of the same general character as before; that is we observe peaks corresponding to maximal contact structures - which are labelled using the notation (m,n) in table I - and decreasing or increasing $D/d$ yields structures with fewer contacts which can in some way be deformed steadily into the maximal contact packing - these are labelled $-(m,n)$ and $+(m,n)$ respectively.

Packings with internal spheres are depicted in Fig (\ref{higher_order}) where in each case the middle diagram shows the entire structure viewed side-on. Although the introduction of inner spheres produces more complicated structures there remains an outer shell of spheres contacting the confining cylinder. The rolled-out pattern made by these outer spheres is shown to the left of each packing. The arrangement of the inner spheres is depicted in the diagram to the right of each packing, whereby the outer layer is transparent and the inner spheres are in yellow.

The first packing with internal spheres is structure 30, or $-(1,5)$, which consists of a basic unit cell containing six spheres. Of these one sphere is not in contact with the cylindrical surface and is found at the centre of the cylinder. The remaining five spheres, which are in contact with the cylinder, form a tilted pentagonal ring around the central sphere (i.e. tilted at an angle with respect to the plane perpendicular to the cylindrical axis). The extended structure can be made by translating the unit cell along the cylinder and rotating it by $\pi$. The result is a chiral packing with an internal chain of non-contacting spheres along the cylindrical axis and an outer layer of touching pentagonal rings. As can be seen, in the corresponding rolled out diagram of the outer layer in Fig (\ref{higher_order}), there is only a single point of contact between outer rings corresponding to a sphere with four surface neighbours. The location of this sphere alternates by an azimuthal angle of $\pi$ between successive layers.

Some of the spheres in the outer layer have only two contacts with other spheres in the outer layer. However, mechanical stability is assured by their contact with the internal chain of spheres and the average number of contacts per sphere for the structure as a whole is above  four.

Increasing $D/d$ reduces the tilt of the pentagonal rings until at $D/d=2.7306$ we have the maximal contact packing  $(1,5)$. Thus in structure 31 the spheres in the internal chain are in contact with each other and, compared with structure 30, there are now a greater number of contacts between successive pentagonal rings. Increasing $D/d$ breaks some of these surface contacts and forces the chain of internal spheres to form a twisted zigzag structure. Thus structure 32, or $+(1,5)$, is chiral and is superseded at $D/d=2.74804$ by a new type of packing.

Structure 33, or $-(1,10)$ consists of a basic unit cell containing eleven spheres. Of these ten touch the cylindrical surface and are arranged as a pair of pentagonal rings stacked on top of each other. The eleventh, internal, sphere is found above the ten surface spheres and is located at the centre of the cylinder. The extended structure can be described as an alternating sequence of surface spheres followed by an internal sphere. However, on average each sphere is only in contact with $40/11=3.636$ other spheres, this number is too low for mechanical stability, which we explain as follows.

As shown by the rolled out diagram of the outer sphere centres, surface spheres from adjacent unit cells are not in contact. Thus the pairs of pentagonal rings are free to rotate, by a certain amount about the cylindrical axis while internal spheres remain fixed in position. Whether structure 33 is achiral or chiral depends on the relative orientation of the surface spheres with respect to each other. No other optimal cylindrical packing yet discovered has this property.

With increasing $D/d$ the pentagonal rings from adjacent unit cells are eventually locked into position at $D/d=2.8211$ to give $(1,10)$. Thus packing 34 is a maximal contact achiral structure, corresponding to a peak in Fig (\ref{volume_fraction}). The spheres in the outer layer form a perfect rhombic $[5,5]$ lattice when rolled out onto the plane and the internal spheres lie along the cylindrical axis. An increase in $D/d$ results in the loss of two contacts (on average) between the surface spheres and the internal spheres, and $+(1,10)$ proceeds downwards in density. As the trend continues we observe a decrease in the separation of neighbouring internal spheres and a modulation in the rhombic surface pattern; this latter symmetry breaking is responsible for structure 35 being a chiral structure.

Packing 36 includes an internal linear chain of non-contacting spheres lying along the central axis of the confining cylinder. These are surrounded by an outer layer of spheres which form a complex chiral structure.   As $D/d$ is increased the spheres in the internal chain are brought ever closer to each other until they make contact at $D/d = 2.8615$, which corresponds to the maximal contact, chiral, structure 37.

A further increase in $D/d$ forces the internal chain of spheres to form a twisted zigzag structure. As a result there is a loss of two types of contacts: the breaking of contacts between spheres in the outer layer (as seen from the corresponding rolled out pattern in Fig (\ref{higher_order})) and a break in contact between spheres in the outer and inner layer. Thus there is a decrease in density as we proceed from structure 37 along 38, i.e. as we increase $D/d$ for $+(2,5)$.

Structure 39 is remarkable in that we find an internal chain of spheres (along the central axis of the cylinder) but the chain is composed of a pair of touching spheres followed by a gap, this is the structure $-(2,13)$. Increasing $D/d$ produces structure 40, a maximal contact chiral packing $(2,13)$, which compared with structure 39 has an increased number of contacts between internal-surface and surface-surface spheres.

This remarkable sequence of structures would have been difficult to imagine in advance. It is likely to be followed by an equally rich scenario, whenever it becomes possible to pursue higher values of $D/d$. The structures reported are new, to our knowledge, except in so far as they correspond to dry foam structures mentioned below in section $10$.

\section{Related Experiments}

Columnar sphere packings appear in a variety of different experimental contexts, such as the packing of $C_{60}$
buckyballs inside carbon nanotubes \cite{Khlobystov:2004} and polystyrene
spheres inside the pores of silicone membranes \cite{Tymczenko:2008}.
The structures that are commonly identified are the straight chain
(structure 1 in
our table I), zigzag (structure 2 or $C_1$), twisted zigzag (structure
3) and structure 5, although more complicated structure have also been
observed, e.g. \cite{Li:2005}.

However, care needs to be taken when directly comparing these experimental
observations
with the simple hard sphere simulations described
here, which might well only
serve as a first guide of what
structures to expect. The minimization of interaction energy may replace the maximization of density as the guiding principle.

For example the polystyrene particles in the
experiments of \cite{Tymczenko:2008} have been charge-stabilised and thus
repel each other. Unlike the situation in our simulations, this results in a preference to sit near the wall of the
pores (i.e. the cylinder wall).

Also the comparison of our simulation results with packings of buckyballs
might be limited. Experiments by \cite{Khlobystov:2004} show that while it
is possible to fill $C_{60}$ into {\em double}-walled nanotubes down to the
ratio $D/d = 1.13$ (resulting in a simple linear chain), it is not possible to
fill {\em single}-walled carbon nanotubes below $D/d = 1.25$ ($D$ is the inner
diameter of the tube in both cases). This reflects the role of the van der
Waals forces between  $C_{60}$ and the confining nanotube walls.

\subsection{Present experiments with ball bearings and bubbles}

However, we have identified two other experimental systems that provide
very good comparison with our simulation data (details to be provided in a
follow-up paper).

The first are
metal spheres of a few millimeter in diameter (the type used in ``ball
bearings'') packed into perspex tubes.
The filled tubes were mechanically agitated over an extended period. At
the end of this ``annealing'' procedure this resulted in clearly
identifiable ordered sphere packings.
These experiments were carried out for a large range of values of D/d (up to
3.15),
and resulted in about 16 different structures, of both the maximal contact
and the line-slip type.

We have also carried out more extensive experiments with equal-volume gas
bubbles (with diameter of a few hundred microns). These are produced
by bubbling gas into surfactant solution with the bubble diameter easily
variable by adjusting the gas flow.

The bubbles are then gathered into vertically placed cylinders.
They maintain a (nearly) spherical
shape, even when in contact, up to a column height of a few millimetres,
corresponding to the capillary length for the surfactant solution in use.

We were able to produce straight chain, zigzag and a large number
of maximal contact structures, and also many structures with internal bubbles, up to and well beyond the point reached by the present simulations. There is also some evidence of the twisted
zigzag structure and line-slip structures. The subsequent paper presenting these experimentally obtained structures will include the use of X-ray tomography to establish their internal configurations.

\subsection{Ordered dry foam structures}

Extensive and detailed results have been published for the various structures that form when
equal volume bubbles with diameter exceeding  a few millimeters (i.e.
larger than the capillary length) are collected in vertical tubes
\cite{Pittet:1995,Tobin:2011}. In this
case, the liquid drains and a \emph{dry}
foam is created - a packing of polyhedral bubbles, with its volume fraction
$\Phi$ approaching unity.

\begin{figure}[tbp]
\includegraphics[width=7cm]{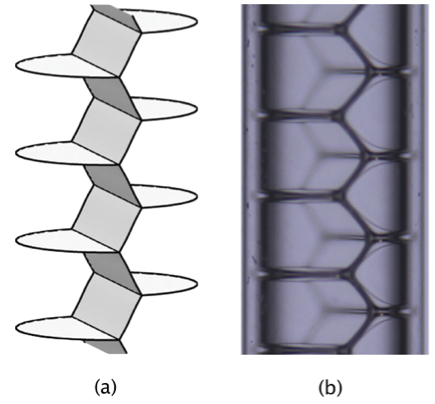}
\caption{In the dry foam equivalent of the zigzag structure
(often called the staircase structure) equal-volume gas bubbles are
separated by thin liquid films. (a) Computer simulation using the Surface
Evolver Software of Ken Brakke \cite{Brakke}. (b) Photograph of bubbles in
a cylinder of about $1cm$ in diameter.}
\label{dry_foam}
\end{figure}

The surface pattern displayed by these ordered foams is hexagonal (apart
from the so-called bamboo structure which is simply an array of parallel
soap films). The phyllotactic notation is thus the obvious choice for their
classification, at least as long as there are no internal bubbles (see for example Fig.
\ref{dry_foam}).

Restricting ourselves for the moment to this case it would appear that {\em all}
ordered dry foam structures have corresponding sphere packings (wet foams), i.e. they can be classified by the same phyllotactic notation. These include the
straight chain (bamboo), and zigzag structure
(called 2-1-1 or staircase structure in
the foam literature), and all of the 10 maximal contact structures,
except structure 25 (5-4-1) which has not yet been observed. {\em None} of the ordered dry foam structures is of the
line-slip type.

Note that since the bubbles are deformable, the average contact number
in the case of dry foams is generally higher than in
hard sphere packings.

Unlike their sphere packing counterparts, the various dry foam structures
are found over ranges of values of $D/d$ (in the foams
literature $d$ refers to the equivalent
sphere diameter of a bubble).  Hysteresis plays a large role in the
standard experimental procedure, leaving to overlapping ranges of stability
for the various structures.

The structure of minimal energy for a
given value of $D/d$ may be determined from computer simulations
\cite{WeairePhelan1996, WeaireBradleyPhelan2000, SaadatfarEtal08, Hutzler:2009, Tobin:2011}. We find these ranges always to be lower than the value
of D/d for the corresponding sphere packing, reflecting their much smaller
volume fractions. It remains for future research to establish the phase diagram (which may be rather complex) that connects these two limiting cases.

There are also observations of over 20 different ordered dry foam
structures with internal bubbles \cite{Pittet:1995}, but the precise internal arrangements
have only been identified for the three simplest cases, with the
aid of Surface Evolver simulations \cite{Brakke,SaadatfarEtal08}.

The dry foam with a surface pattern equivalent to structure 28 ($C_5$)
has six bubbles in the periodic unit cell, i.e. one more than $C_5$.
A pentagonal dodecahedron in the centre of the cylinder is surrounded by a
ring of five bubbles in contact with the cylinder surface. The dodecahedra
of neighbouring unit cells are in contact.

There is also a foam structure with a so-called Kelvin cell
(tetracaidecahedron) in the centre, surrounded by 6 bubbles in contact with
the surface. Again the internal bubbles of neighbouring unit cells are in
contact with each other.

The third identified structure with
internal bubbles consists of a
a total of 13 bubbles in the unit cell, 12 touching the surface,
and a bubble with 12 pentagons and  3
hexagons (Goldberg-3) in the centre \cite{SaadatfarEtal08}.
Interestingly in this case the internal bubbles of neighbouring unit cells
are not in contact, similar to the also achiral structure 34.
The surface pattern of this foam structure consists of an arrangement of
bi-disperse hexagons.

\section{Conclusions}
Our simulations have provided detailed results for 40 distinct columnar crystals, of which many are new. They fall into three categories: the very simplest cases, where the sphere diameter is of the same order as that of the cylinder, a wide range of further "phyllotactic" structures, which may be understood as related to surface packings of disks, and structures that incorporate internal spheres.

Corresponding experimental observations are now available and will be reported in a second paper. Experiments can already be pushed to much larger cylinder diameters, providing further insights and challenges to simulation and interpretation.

Such a further development should motivate a reassessment of the simulation methods to be used; we make no strong claims for the efficiency of that used here. We have necessarily been cautious in using it, stipulating high degrees of convergence and undertaking many repeated runs in search of the optimum, to increase confidence in our conclusions. In parallel studies, Chan \cite{chan:2011} used a sequential deposition method that was highly expeditious in determining optimal structures within the restriction of that procedure, whose effect could not be known a priori. Suitably adapted, it was able to reproduce the structures reported here, up to at least $D/d = 2.7013$. Future work will include an extension of this deposition approach to higher values of $D/d$.

Finally we note some possible extensions to this work. An example is the study of disordered hard sphere packings in a cylinder; for such problems the definition of the volume fraction recently given by Chan \cite{chan:2011} may prove useful. A simple problem that may be of significant practical interest is to find the densest packing of hard spheres in a cylinder which is capped on either side by hard walls (i.e. packing in a cylindrical box). Even richer possibilities are offered by packing hard spheres in a cylinder that is capped on both ends by spheres held in a fixed position. Indeed, hard sphere packings that are bounded by templated surfaces on all sides are of significant current interest and are capable of realising a range of different morphologies (including the Weaire-Phelan structure)  \cite{Gabbrielli:2012}.

\section{Acknowledgments}

This research was supported by Science Foundation Ireland (08-RFP-MTR1083) and
European Space Agency (MAP AO-99-108:C14914/02/NL/SH and
AO-99-075:C14308/00/NL/SH). H. K. C. would like to acknowledge support from the Irish Research Council for Science, Engineering and Technology (EMPOWER Fellowship).

\section{Appendix A: Simulation technique}
\label{sec:numerical_simulations}

\subsection{Energy function}

The simulation is addressed to a cylindrically shaped cell of length $L$ and diameter $D$. Contained within this space are $N$ points which represent the centres of $N$ spheres, each of diameter  $d$.  If a pair of spheres is sufficiently close that they overlap we account for this using a pairwise potential as described below. A similar overlap potential is used to prevent the spheres from escaping the simulation cell in the radial direction. The final structures will be the densest that we find which have zero overlap energy.

In addition, we impose  \textit{twisted} periodic boundary conditions on the top and bottom of the cylindrical cell as described in the main text, with a twist angle $\alpha$.

We model the overlap potential between spheres using a Hookean, or ``spring-like'', pairwise interaction between the $i$th and $j$th spheres, which have their centres at  ${\bf r}_i=(r_i, \theta_i, z_i)$ and ${\bf r}_j=(r_j, \theta_j, z_j)$, the interaction energy between spheres is then given by,
\begin{equation}
E^S_{ij}
 =
\left\{
\begin{array}{l l}
\frac{1}{2}(r_{ij}-d)^2 & \quad \mbox{if $r_{ij}\leq d$}\\
0 & \quad \mbox{if $r_{ij} >d$}\\
\end{array}
\right.
\end{equation}
where $r_{ij}=|{\bf r}_i-{\bf r}_j|$ is the distance between the centres of the spheres. Note the interaction energy  falls to zero when there is no overlap between the spheres.

The interaction energy between the $i$th sphere and the boundary is given by,
\begin{equation}
E^B_{i}
=
\left\{
\begin{array}{l l}
\frac{1}{2}(r_{iB}-d/2)^2 & \quad \mbox{if $r_{iB}\leq d/2$}\\
0 & \quad \mbox{if $r_{iB} >d/2$}\\
\end{array}
\right.
\end{equation}
where $r_{iB}=|D/2-r_i|$.

The \textit{twisted} periodic boundary conditions on the ends of the cylinder are incorporated as follows: The $i$th particle in the simulation cell has an image at the top and bottom of the simulation cell, the coordinates of these images are given by ${\bf r}_i^+=(r_i, \theta_i+\alpha, z_i+L)$ and ${\bf r}_i^-=(r_i, \theta_i-\alpha, z_i-L)$, respectively, where $L$ is the length of the cylinder and $\alpha$ is the twist angle.

Thus the total energy of the system is given by the sum of the sphere-sphere, sphere-boundary and sphere-image interactions.

\subsection{Numerical Method}

For a tube of diameter $D$ we wish to find the unit cell, composed of $N$ spheres, which when rotated and stacked along the tube has the highest volume fraction. In this section we describe the simulation protocol used to achieve this.

For a given $D$ and $d$ we assign initial starting positions to the $N$ spheres and an initial value to the twist angle $\alpha$ by using a random number generator. A small initial value for the cylinder length $L$ is chosen to insure overlap between the spheres.

Keeping $D$ and $L$ fixed, we search for the lowest energy arrangement for the $N$ spheres by varying their coordinates and the twist angle. This is done using the standard Metropolis simulated annealing algorithm, where for a cluster of $N$ spheres the algorithm was run with typically $N\times(5\times10^6)$ Monte Carlos steps. The temperature of the simulation was decreased linearly. The average displacement of the spheres at each temperature step was chosen by an automatic process to give an acceptance probability of $0.5\pm 0.01$. The results of the simulated annealing are then put through a conjugate gradient routine to ensure that a local minimum has been reached.

The whole process is repeated many times, using a new randomly generated initial configuration, to give confidence that the lowest energy state has been found. From this ensemble we take the lowest energy configuration as the final state for that particular run.

After this first run we perform a subsequent run with a slightly longer cylinder. Since the spheres have more room the energy of the final state is lower compared to the final state of the previous run. Using a divide and conquer approach we are able to establish the cell length at which the energy per sphere in the system falls to zero, in practice this means the value of the energy is within a critical bound which is set to be E/N=$1\times10^{-8} \pm 5\times10^{-9}$. At this point the spheres have a small overlap corresponding to a five decimal place accuracy in the volume fraction (this is deduced by comparing simulation with the analytically derived volume fraction for the $C_N$ circle packing structures).

From this final result we compute the volume fraction which is defined as,
\begin{equation}
\Phi(N,D)
=
\frac{NV_s}{V_c}
\end{equation}
where $V_s=(4/3)\pi (d/2)^3$ is the volume of one of the hard spheres and $V_c=\pi (D/2)^2 L$ is the volume of the simulation cell. Note that the volume fraction depends on both $D$ and $N$ since different values of $N$ yield unit cells with different structures. Thus the whole procedure is repeated for series of different values of $N=1,2,3...$, typically up $N=15$ for the structures without internal spheres, before accepting the one which gives the largest volume fraction, as shown in Table \ref{exp:vf_table}.

\section{Appendix B: Continuity of the volume fraction}

We demonstrate that there can be no sudden finite discontinuities in the curve describing the maximum volume fraction (density) as $D/d$ is varied. This follows from lower and upper bounds, as described below, which limit the variation in volume fraction in the neighbourhood of any point $D_0/d_0$. We will see that square root singularities observed in the numerical results for the volume fraction are due to the arguments given below.

\subsection{Lower bound}

For increasing $D/d$ an obvious variational argument bounds the density below. As $D/d$ is increased from $D_0/d_0$, the cylinder expands radially. We may take the structure which has maximum density at $D_0/d_0$ as a trial structure for $D/d> D_0/d_0$. The volume fraction of the structure, at $D_0/d_0$ is,
\begin{equation}
\Phi(D_0/d_0)= \frac{V_s}{\pi (D_0/2)^2 \bar{L}},
\label{eq:vf_d0}
\end{equation}
and that of the trial structure is,
\begin{equation}
\Phi^{\mathrm{T}}(D/d)= \frac{V_s}{\pi (D/2)^2 \bar{L}},
\end{equation}
where $V_s=V_s(d)$ is the volume of a sphere of diameter $d$ and $\bar{L}$ is the average separation between spheres in the $\widehat{\bf z}$ direction. Since $\Phi^{\mathrm{T}}(D_0/d_0)=\Phi(D_0/d_0)$ it follows that,
\begin{equation}
\Phi^{\mathrm{T}}(D/d)>\frac{D_0^2}{D^2}\Phi(D_0/d_0)
\end{equation}
providing a lower bound for $D>D_0$.

\subsection{Upper bound}

For decreasing $D/d$, a more subtle argument bounds the variation of density below by a square-root function. Decreasing $D/d$ forces the spheres to move radially inwards to avoid contact with the cylindrical boundary. The resulting overlap between spheres can be eliminated by displacement of their centres parallel to the cylinder axis.

Let us index the sphere centres in ascending height using the index $j$ - so that $z_j> z_{j-1}$ for $j>j-1$. Given the structure at $D_0/d_0$ which has a maximum density, then an extreme case is one where successive spheres all have the same height. Consider in this case a pair of contacting spheres with separation $d$. Reducing the cylinder diameter by a factor of $X$, so that the new diameter is $D=D_0X$, will force an overlap so that the separation between sphere centres is now $dX$. The overlap can be removed by moving one of the spheres vertically a distance $\Delta$, so that their separation is once again $d$. It follows that,
$d^2=\Delta^2 - (dX)^2$. Thus a constant $C$ can be chosen so that the following choice is sufficient,
\begin{equation}
\Delta=C\sqrt{D^2_0-D^2}.
\end{equation}
In order to eliminate the overlap, the sphere centres are shifted to a new height $z'_{j}=z_j+j\Delta$. The resulting trial structure has the volume fraction,
\begin{equation}
\Phi^{T}=\frac{V_s}{\pi (D/2)^2 (\bar{L} + \Delta)}
\end{equation}
which when combined with Eq. ($\!\!$~\ref{eq:vf_d0}) results in the following bound
\begin{equation}
\Phi^{T}(D/d)<\frac{D_0^2}{D^2}\frac{\bar{L}}{\bar{L}+\Delta}\Phi(D_0/d_0).
\end{equation}
Hence we again arrive at a (lower) bound for $\Phi^{T}(D/d)$ which goes continuously to $\Phi(D_0/d_0)$ as $D/d\rightarrow D_0/d_0$ but in this case with a square root form.

\newpage

\bibliographystyle{nonspacebib}

\end{document}